\documentclass[%
 reprint, 
superscriptaddress,
 amsmath,amssymb,
 aps,
pra,
]{revtex4-2}
\usepackage{xcolor}
\usepackage{graphicx}
\usepackage{dcolumn}
\usepackage{bm}
\usepackage{hyperref}
\hypersetup{
    colorlinks=true, 
    linkcolor=black, 
    citecolor=black, 
    urlcolor=black 
}
\usepackage{ulem}
\usepackage[T1]{fontenc}

\begin{document}

\title{On-Demand Microwave Single-Photon Source Based on Tantalum Thin Film}
\author{Ying Hu}
\affiliation{Key Laboratory of Low-Dimensional Quantum Structures and Quantum Control
of Ministry of Education, Department of Physics and Synergetic Innovation
Center for Quantum Effects and Applications, Hunan Normal University,
Changsha 410081, China.}
\author{Sheng-Yong Li}
\affiliation{Department of Automation, Tsinghua University, Beijing 100084, China.
}
\author{En-Qi Chen}
\affiliation{Key Laboratory of Low-Dimensional Quantum Structures and Quantum Control
of Ministry of Education, Department of Physics and Synergetic Innovation
Center for Quantum Effects and Applications, Hunan Normal University,
Changsha 410081, China.}
\author{Jing Zhang}
\affiliation{School of Automation
Science and Engineering, Xi’an Jiaotong University, Xi’an, 710049, China.}
\affiliation{MOE Key Lab for Intelligent Networks and Network Security, Xi’an Jiaotong University,
Xi’an, 710049, China.}
\author{Yu-xi Liu}
\affiliation{School of Integrated Circuits, Tsinghua University, Beijing 100084, China.}
\author{Jia-Gui Feng}
\email{fengjiagui@quantumsc.cn}
\affiliation{Gusu Laboratory of Materials, Suzhou 215123, China.}
\affiliation{Suzhou Institute of Nano-Tech and Nano-Bionics, CAS, Suzhou 215123,
China.}
\author{Zhihui Peng}
\email{zhihui.peng@hunnu.edu.cn}
\affiliation{Key Laboratory of Low-Dimensional Quantum Structures and Quantum Control
of Ministry of Education, Department of Physics and Synergetic Innovation
Center for Quantum Effects and Applications, Hunan Normal University,
Changsha 410081, China.}
\affiliation{Hefei National Laboratory, Hefei 230088, China.}

\date{\today}

\begin{abstract}
Single-photon sources are crucial for quantum information technologies. Here, we demonstrate a microwave single-photon source fabricated using a tantalum-based thin film, whose favorable material properties enable high-quality and stable photon emission. The antibunching behavior of the emitted radiation is revealed by second-order correlation measurements. Furthermore, traveling-wave parametric amplifiers are used as the pre-amplifier in the detection chains, we substantially improve the signal-to-noise ratio and thereby greatly reduce the acquisition time required for second-order correlation measurements. These results demonstrate the viability of tantalum-based superconducting devices as reliable platforms for microwave quantum photonics.
\end{abstract}

\maketitle


\section{Introduction}

On-demand single-photon sources are key resources for quantum information science and technology~\cite{Cirac_PRL_1997,Zoller_EPJD_2005,Kimble_nature_2008,Couteau_NRP_2023}. While single-photon sources at optical regime have already become standard tools in quantum communication and photonic quantum computation, their counterparts in the microwave regime play an equally central role in superconducting quantum computing and circuit quantum electrodynamics (cQED) architectures~\cite{Kjaergaard_ARCMP_2020,Blais_RevModPhys_2021}. Propagating microwave single photons provide a natural interface between stationary superconducting qubits and itinerant bosonic modes, enabling remote entanglement distribution~\cite{Kurpiers_Nature_2018,Axline_NatPhy_2018}, quantum communication between quantum nodes~\cite{Zhong_NatPhys_2019,Zhong_Nature_2021,Qiu_SciBull_2025}, and quantum repeaters in the microwave domain~\cite{Jiang_nphQI_2021,Azuma_RMP_2023}. Moreover, the realization and precise characterization of high-fidelity microwave single-photon sources has become a primary objective in the development of large-scale superconducting quantum processors and distributed quantum networks.

Over the past decade, a variety of schemes for generating propagating microwave single photons have been proposed and experimentally demonstrated, typically using coplanar waveguide resonators or superconducting qubits coupled to one-dimensional transmission lines~\cite{Bozyigit_NatPhys_2011,Hoi_PRL_2012,Lang_NatPhys_2013,Peng_NatCommun_2016,Zhou_PhysRevAppl_2020,Lu_NpjQuantumInf_2021}. Despite these advances, experiments in the microwave domain face several intrinsic challenges that are qualitatively different from those encountered in the optical regime. First, the energy of a microwave photon is several orders of magnitude smaller than that of an optical photon, which renders microwave signals extremely susceptible to thermal noise even at low temperatures. Second, practical, high-efficiency, number-resolving microwave single-photon detectors are still under active development and are not yet as mature or widely available as their optical counterparts~\cite{Lubsanov_SST_2022}. As a consequence, most current experiments rely on linear amplification followed by heterodyne (or homodyne) detection of the field quadratures~\cite{Bozyigit_NatPhys_2011,Hoi_PRL_2012,Lang_NatPhys_2013,Peng_NatCommun_2016,Zhou_PhysRevAppl_2020,Lu_NpjQuantumInf_2021}. The single-photon character of the emitted field and its higher-order correlation functions must then be inferred only indirectly, through statistical analysis of the recorded quadrature time traces together with  deconvolution of the entire measurement chain, which typically results in very long integration times. {Meanwhile, the rapid development of parametric amplification has  reshaped microwave quantum measurements. Early Josephson parametric amplifiers~\cite{CastellanosBeltran_APL_2007,Yamamoto_APL_2008,Aumentado_IEEE_2020}, and more recently traveling-wave parametric amplifiers (TWPAs)~\cite{HoEom_NP_2012,Brien_PhysRevLett_2014,Bockstiegel_JLTP_2014,Macklin_Science_2015,Esposito_APL_2021}, can now routinely provide near–quantum-limited noise performance over multi-gigahertz bandwidths with high saturation power.}

 From a materials-science perspective, the exploration of low-loss superconducting thin films and interfaces has led to substantial improvements in the coherence of superconducting resonators and qubits, as well as a marked reduction of dielectric loss~\cite{Place_NC_2021}. $\alpha$-Ta (110) thin films has been introduced as a novel material platform for the development of superconducting quantum circuits~\cite{Ding_JVSTB_2024}. This has led to an increase in the coherence time of Transmon qubits to $300\,\mu$s, which was later optimized to $500\,\mu$s~\cite{Place_NC_2021,Wang_npjQI_2022}. A superconducting quantum chip with over 100 qubits, based on $\alpha$-Ta (110) film, has also been developed, with a median coherence time exceeding $100\,\mu$s~\cite{Zhou_JJAP_2023}. The significant improvement in coherence times using $\alpha$-Ta (110) film has been attributed to the formation of a dense amorphous $\mathrm{Ta_2O_5}$ passivization layer with low microwave loss on the surface of the thin film during the piranha solution treatment process. This $\mathrm{Ta_2O_5}$ is highly stable and exhibits minimal aging effects on the intrinsic quality factor of the superconducting resonator in ambient conditions over extended periods. {The employment of artificial atoms based on tantalum thin film as microwave single-photon sources offers an alternative platform for implementing high-performance microwave single-photon sources.}

In this work, we fabricate a microwave single-photon source based on tantalum thin films and demonstrate sub-Poissonian photon statistics and antibunched emission via correlation measurements. Especially, TWPAs are used as the low-noise pre-amplifiers in the detection chain for pulsed single-photon detection, we enhance the signal-to-noise ratio and significantly reduce the averaging time required for second-order correlation measurements.

The paper is structured as follows: In Sec. II, we describe the fabrication process of the samples. Sec. III provides an overview of the experimental setup. In Sec. IV, we present frequency-domain measurements that characterize the device performance. Finally, Sec. V reports time-domain and correlation measurements, which validate the single-photon character of the source.

\section{The device fabrication}

\begin{figure}[b]
\includegraphics[width=0.49\textwidth]{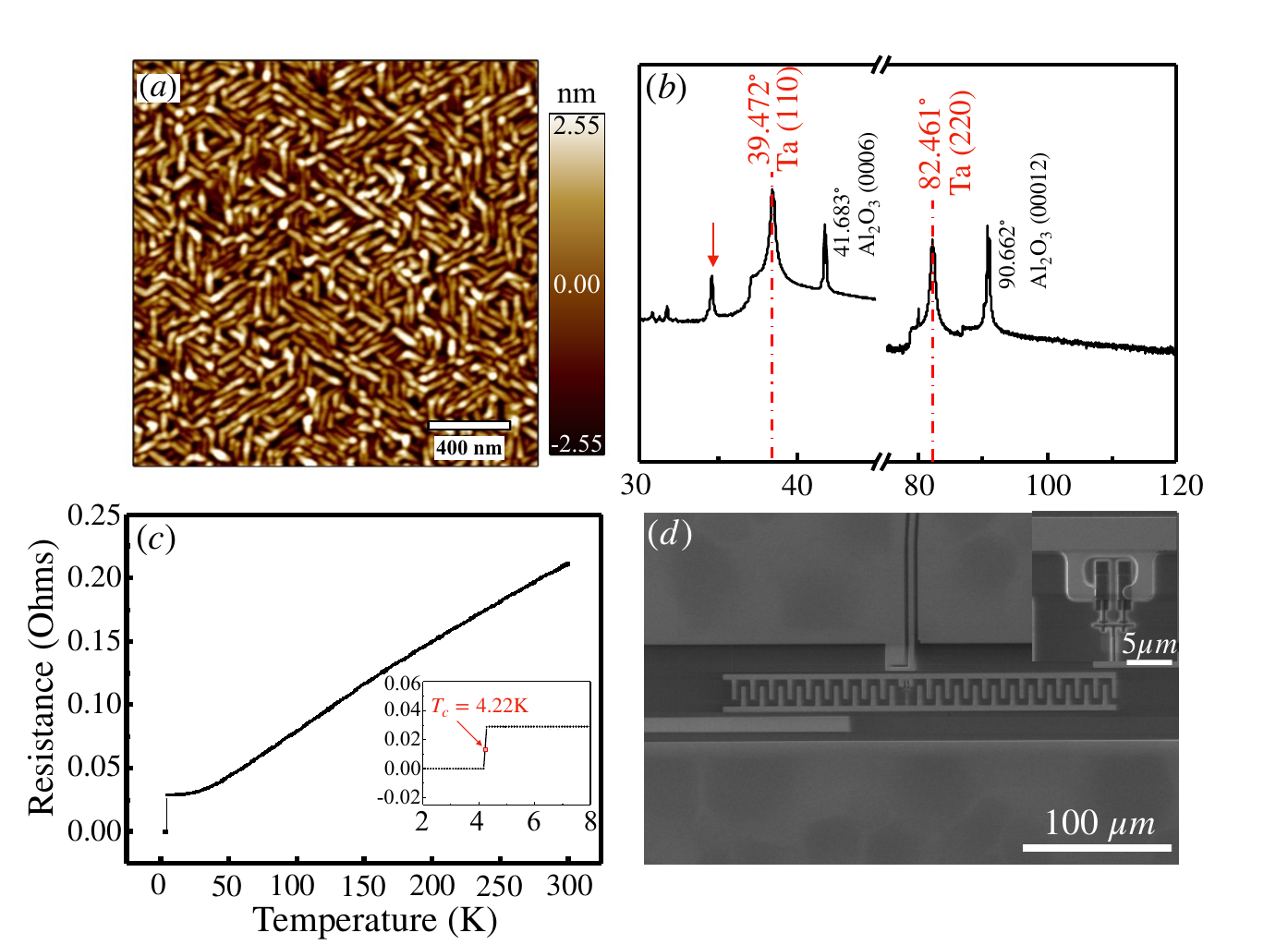}\caption{\label{fig:SampleFab.} The thin film and device characterization: (a) AFM image of surface morphology.
(b) XRD characterization of the film with red dashed lines indicating
the (110) and (220) peak positions, the $\beta$-Ta (002) peak is
marked by red arrow. (c) Measurement curve of film resistance vs
temperature, with inset showing the superconducting transition temperature.
(d) SEM characterization of the key components of the microwave single-photon
source. The inset is a zoom-in image, which shows the SQUID of the
qubit.
}
\end{figure}

Single photon sources are created using high-grade tantalum films. $\alpha$-Ta (110) films, with a thickness of 200 nm, are deposited on 2-inch C-plane sapphire substrates via DC magnetron sputtering within a high vacuum chamber. Before film deposition, the substrate is subjected to a thermal cleanse inside the sputtering chamber at $700^{\circ}\mathrm{C}$ for $30$ minutes, followed by a gradual cooling down to $400^{\circ}\mathrm{C}$ at a rate of $30^{\circ}\mathrm{C}$ per minute. Throughout the deposition process, the substrate is kept at a constant temperature of $400^{\circ}\mathrm{C}$. Ar gas, flowing at a rate of $30$ sccm, is continuously introduced into the chamber. The sputtering pressure is maintained at a constant $15$ mTorr, while the DC sputtering power is set to $200$ W. As illustrated in   Fig. 2(a), the atomic force microscope image reveals that the grains of Ta are elongated, possessing tetragonal symmetry, and are densely packed, with a surface roughness (Rq) of 1.25 nm in an area of 2 $\mu$m $\times$ 2 $\mu$m. The $x-$ray diffraction measurements (presented in Fig. 2(b) identify the main peaks of Ta (110), Ta (220), $\mathrm{Al_{2}O_{3}}$ (0006), and $\mathrm{Al_{2}O_{3}}$(00012), with a small peak corresponding to the $\beta$-Ta film indicated by a red triangle. The Scherrer analysis reveals crystallite sizes of approximately 100 nm. 

The temperature-dependent resistance of the Ta film is depicted in 
Fig. 2(c). The figure reveals that the superconducting transition
temperature of the film is 4.18 K, with a transition temperature range
of 0.1 K (as shown in the inset of Fig. 2(c). We calculate the residual
resistance ratio (RRR) as the ratio of the measured resistance between
$300$ K and $10$ K, which yields a value as high as $7.3$. This
value surpasses the RRR of the Ta film used in a qubit, which has
a coherence time of $500$ $\mu$s~\cite{Wang_npjQI_2022}. The RRR of the film is
significantly influenced by several factors, such as film quality,
thickness, surface, and interfaces between the film and the substrate,
all of which are associated with microwave loss in superconducting
microwave devices. The RRR has long been employed as a metric to evaluate
the quality of superconducting films for the construction of superconducting
radiofrequency cavities ~\cite{Russo_SST_2005,Valderrama_AIPCP_2012}. Recent research indicates that
RRR can also serve as a proxy for assessing the performance of superconducting
qubits~\cite{Premkumar_CM_2021}. Thus, the elevated RRR value suggests that the high
quality of a single microwave source can be realized by utilizing
this type of $\alpha$-Ta (110) film as the superconducting circuit
material.

The key components of the microwave single-photon source, as depicted
in Fig. 2(d), are fabricated through the following process: sapphire
wafers with a 200 nm Ta film are subjected to sonication in acetone
and isopropyl alcohol for 5 minutes each, followed by rinsing with
DI water. To enhance the device's stability during air exposure, a
20-minute cleaning with a piranha solution is performed ~\cite{Ding_JVSTB_2024}. The
patterns of the capacitors, resonators, and drive lines are then defined
by photolithography. To remove the unwanted regions of the Ta film,
a wet etching method is employed, primarily to mitigate the microwave
loss resulting from substrate damage ~\cite{Zhou_JJAP_2023}. Finally, the $\mathrm{Al/AlO_{x}/Al}$
Josephson junction is fabricated using a standard method, as illustrated
in ~\cite{Zheng_SR_2023}.

\section{DEVICE AND EXPERIMENT SETUP}

\begin{figure}[b]
\includegraphics[width=0.45\textwidth]{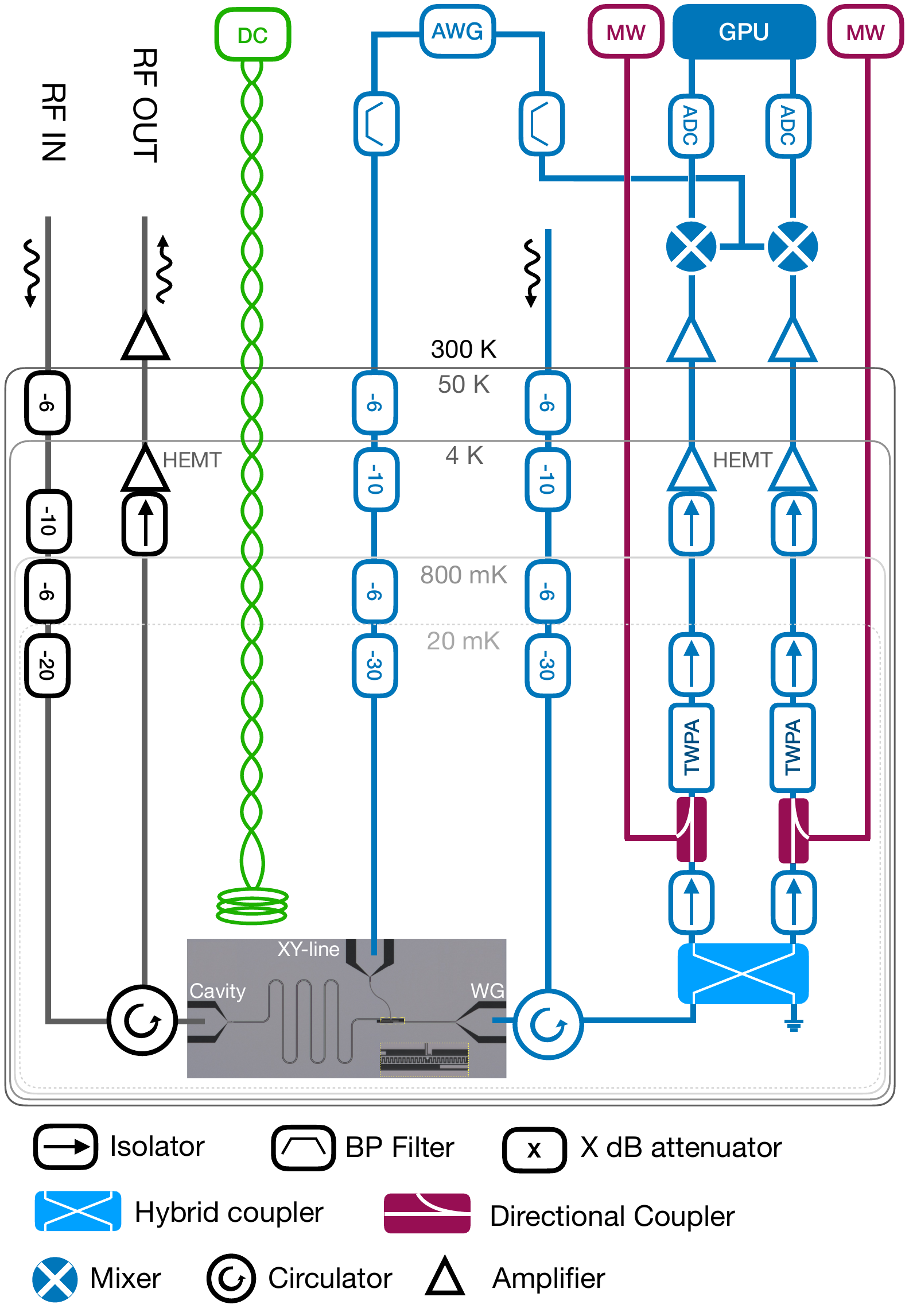}\caption{\label{fig:exp_setup} Schematic diagram of the experimental setup, illustrating both the cryogenic and room-temperature components employed for frequency-domain and time-domain measurements. The cryogenic stage hosts the superconducting quantum device together with the associated input attenuation, DC flux-bias lines, and amplification chain. The room-temperature stage includes the microwave signal sources, DC sources, arbitrary waveform generators, analog-to-digital converters, and the signal-processing unit used for control and measurement.}
\end{figure}

We use tantalum thin-film technology to fabricate the single-photon source described above, which consists of a transmon qubit that is capacitively and strongly coupled simultaneously to a 1D coplanar-waveguide resonator and a transmission waveguide~\cite{Koch_PRA_2007,You_PRB_2007,Bozyigit_NatPhys_2011,Hoi_PRL_2012,Peng_NatCommun_2016,Zhou_PhysRevAppl_2020}. A key advantage of this design is its ability to emit single-photon signal with tunable frequencies.
The qubit's pump signal is configured with a dedicated control-XY
line. {The transmon qubit consists of a dc superconducting quantum interference device  in parallel with a large capacitor, and its transition frequency can be tuned by an externally applied flux bias}. 
In this experiment, the qubit is characterized by a maximum Josephson energy of $E_{\mathrm{J,max}}/2\pi \approx 39.03~\mathrm{GHz}$ and a charging energy of $E_{C}/2\pi \approx 400~\mathrm{MHz}$, extracted from fits to the qubit spectroscopy.
Owing to its coupling to the transmission waveguide, the energy-relaxation time is approximately $60~\mathrm{ns}$ at the working frequency of $8.886~\mathrm{GHz}$.
The qubit transition frequency can be tuned by a quasi-static magnetic field generated by a superconducting coil wound around the sample box, and the initial states are prepared by applying nanosecond-timescale microwave pulses through the control-XY line.

Our experiment is carried out in a dilution refrigerator, with the sample cooled to a base temperature of about $18$ mK and magnetically shielded by two layers of mu-metal and an additional aluminum shield. {Figure \ref{fig:exp_setup} depicts the measurement scheme. Attenuators are distributed across the temperature stages of the dilution refrigerator to suppress thermal noise originating from the higher-temperature stages, thereby thereby preventing unwanted excitation of the qubit.} A
$4-12$ GHz circulator placed on the waveguide and cavity ports allows
measurement of the reflection, which can be used to directly characterize
the performance of the single-photon source~\cite{Peng_NatCommun_2016,Zhou_PhysRevAppl_2020}.
The signal emitted or scattered from the qubit is split into two channels
by a $1-12$ GHz hybrid coupler, forming a Hanbury Brown-Twiss-type (HBT)
setup to measure the time-dependent emission dynamics and correlation
functions~\cite{Brown_Nature_1956,Gabelli_PRL_2004}. In each detection channel, we employ a commercial TWPA (QuantumCTek Co., Ltd.)   as the pre-amplifier to enhance the signal-to-noise ratio (SNR), providing an order-of-magnitude improvement compared with a detection chain using only a HEMT and thereby significantly reducing the required averaging time. To ensure the proper operation of the TWPA, we place a $40$
dB isolator at the input port, and a $20$ dB isolator at the output port, with
a directional coupler providing the driving signal~\cite{HoEom_NP_2012,Brien_PhysRevLett_2014,Bockstiegel_JLTP_2014,Macklin_Science_2015,Esposito_APL_2021}.
An additional isolator is placed in front of the High Electron Mobility
Transistor (HEMT) to protect against back action from cryogenic amplifiers
at the 4K stage. The total gain of the output line is approximately $75$ dB, including about $38$ dB from the HEMT and $37$ dB from room-temperature amplifiers.

We employ heterodyne detection to perform time-domain and correlation measurements, facilitated by a linear amplification chain comprising amplifiers at various temperature stages. As the microwave single-photon detector is still under development, the verification of emitted single-photon signals is conducted through correlation measurements~\cite{Bozyigit_NatPhys_2011,Hoi_PRL_2012,Lang_NatPhys_2013,daSilva_PhysRevA_2010,Peng_NatCommun_2016,Zhou_PhysRevAppl_2020,Inomata_NatCommun_2016,Besse_PRX_2018,Kono_NatPhys_2018}.

\begin{figure*}
\includegraphics[width=0.7\textwidth]{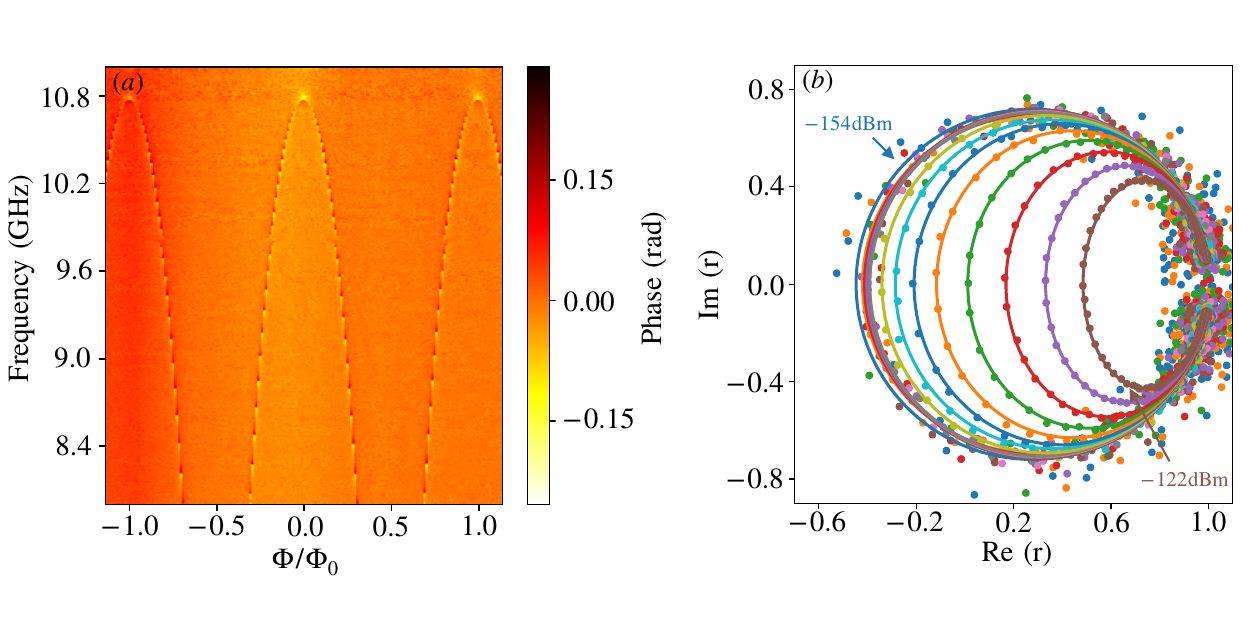}
\caption{\label{fig:energy spectrum}Qubit response under continuous-wave probing.
(a) Reflection spectrum of the transmon qubit as a function of the applied flux bias. The “sweet-spot’’ transition frequency vanishes at integer multiples of the magnetic flux quantum $\Phi_{0}$ threading the SQUID loop.
(b) Complex reflection coefficient $r$ of the waveguide measured at the qubit transition frequency $\omega_{01}/2\pi = 8.887~\mathrm{GHz}$, represented in the real–imaginary plane. Experimental data (dots) are normalized to the background response when $\omega_{01}$ is detuned far from the probing frequency. The probe power is varied from $-154$ to $-122~\mathrm{dBm}$ in steps of $2~\mathrm{dB}$.}
\end{figure*}

The drive-qubit pulse signal is directly emitted from $10$ giga samples-per-second
arbitrary waveform generator (AWG). The emission signal
is down-converted from the qubit frequency to $50$ MHz and digitized
using a two-channel analog-to-digital converter (ADC) with a sample rate of 200 MHz and a time resolution of 5 ns. 
We use a combination of multi-core Central Processing Unit (CPU) and high-performance 
Graphics Processing Unit (GPU)  to rapidly process data streams in real time, obtaining the corresponding
expectation value and correlations efficiently.


\section{SPECTRUM AND EMISSION EFFICIENCY}

We characterize the single-photon source by measuring the reflection coefficient from the waveguide port~\cite{Peng_NatCommun_2016,Zhou_PhysRevAppl_2020}. The cavity mode is coupled to the qubit with a resonance frequency of $\omega_{c}/2\pi = 6.751~\text{GHz}$. Figure~\ref{fig:energy spectrum} shows a two-dimensional $S_{21}$ spectrum as a function of probe frequency and flux bias, demonstrating that the qubit transition frequency, and hence the source frequency, can be tuned by the dc coil up to $10.8~\text{GHz}$. When the probe frequency is resonant with the qubit transition, the reflection spectrum exhibits a pronounced dip.

Next, we characterize the emission efficiency of the qubit at $\omega_{01}/2\pi = 8.887~\text{GHz}$ by measuring the reflection response for drive powers in the range from $-154$~dBm to $-122$~dBm, in order to evaluate its coupling to the waveguide. The radiative decay rate into the waveguide, $\Gamma_{1}^{e}$, contributes to the total energy-relaxation rate $\Gamma_{1} = \Gamma_{1}^{e} + \Gamma_{1}^{c} + \Gamma_{1}^{n}$, where $\Gamma_{1}^{c}$ is the relaxation rate through the control XY line and $\Gamma_{1}^{n}$ represents non-radiative loss channels~\cite{Peng_NatCommun_2016,Zhou_PhysRevAppl_2020,Lu_NpjQuantumInf_2021}. In our device design, the relative coupling strengths of the control line and the waveguide port are chosen such that the majority of the energy is emitted into the waveguide. Under weak driving conditions $(\Omega \ll \Gamma_{1}, \Gamma_{2})$, the reflection coefficient can be expressed as:
\begin{equation}
r=1-\frac{\Gamma_{1}^{e}}{\Gamma_{2}}\frac{1}{1-i\delta\omega/\Gamma_{2}}
\end{equation}
where $\Gamma_{2}=\Gamma_{1}/2+\Gamma_{\phi}$ is the dephasing rate~\cite{Astafiev_Science_2010,Peng_NatCommun_2016,Zhou_PhysRevAppl_2020}.
{Figure~3(b) shows the reflection coefficient $r$ plotted in the complex plane, and a fit to the data yields the decay parameters $\Gamma_{1}^{e}/2\pi = 2.65~\mathrm{MHz}$ and $\Gamma_{2}/2\pi = 1.85~\mathrm{MHz}$.}
The single-photon source efficiency is defined as $\eta = \Gamma_{1}^{e}/\Gamma_{1} \approx \Gamma_{1}^{e}/2\Gamma_{2} = 71.6\%$. As the drive power increases, the qubit's transition gradually saturates,
causing the reflection curve to change its form from circular to oval.
This change reflects the transition from linear weak-driving regime
up to the non-linear strong-driving regime. In the subsequent sections,
we will use short pulse signal instead of continuous signals to improve
the manipulation efficiency.

\section{Rabi Oscillation and Correlation Functions}

\begin{figure*}
\includegraphics[width=0.75\textwidth]{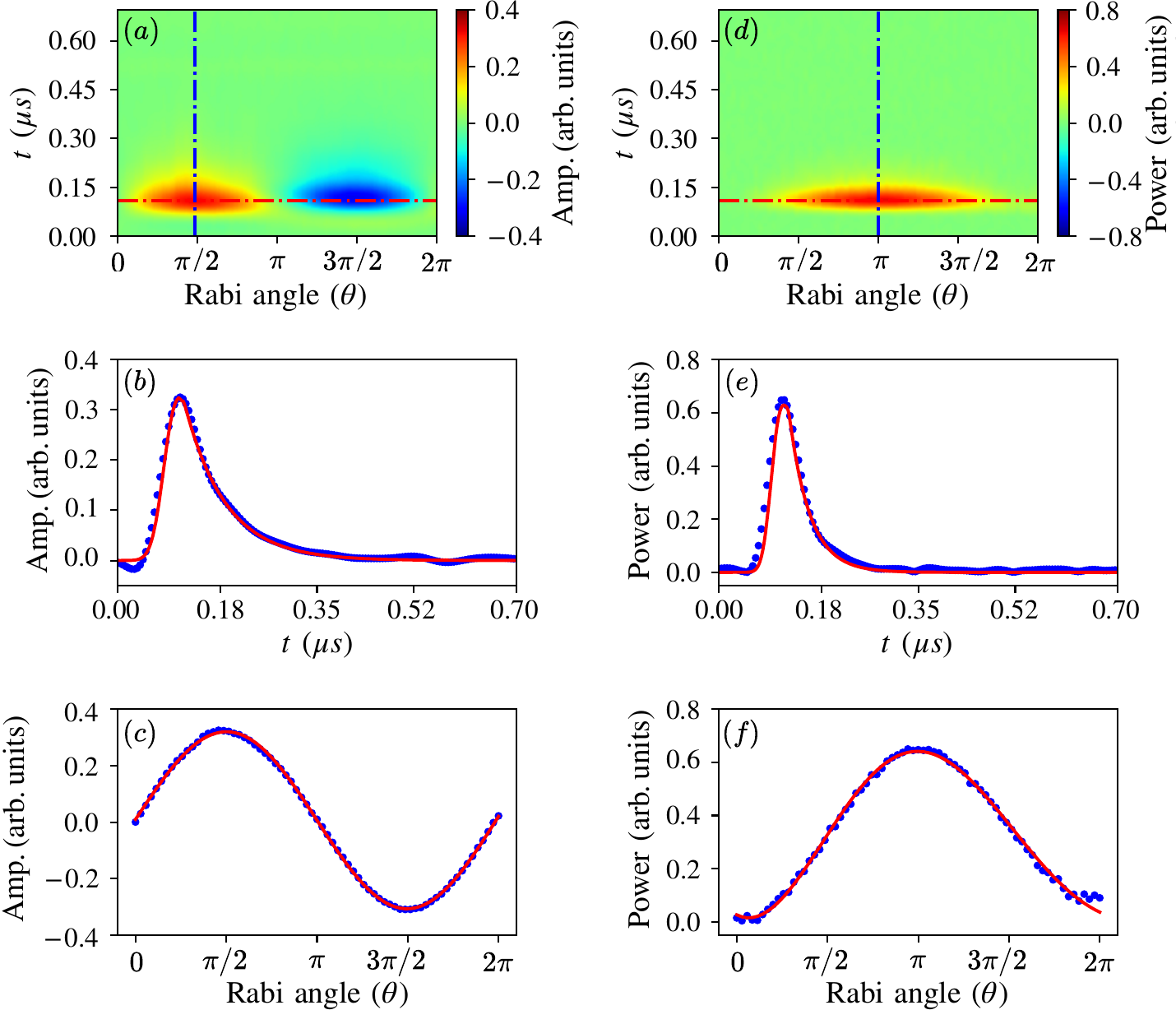}

\caption{\label{fig:Rabi osillation}Rabi oscillation measurement of the emitted field.
(a) Time-resolved quadrature amplitude of the emission field detected at one output port of the beam splitter as a function of the qubit preparation Rabi angle $\theta_{r}$.
(b) A representative quadrature-amplitude trace at $\theta_{r} = \pi/2$, corresponding to the state $(\vert 0\rangle + \vert 1\rangle)/\sqrt{2}$ [blue dash–dotted line in (a)].
(c) Maximum quadrature amplitude extracted from each trace in (a) as a function of $\theta_{r}$ [red dash–dotted line in (a)].
(d) Time dependence of the cross power between the two output channels for the same set of Rabi angles $\theta_{r}$ as in (a).
(e) A representative cross-power trace at $\theta_{r} = \pi$, corresponding to the state $\vert 1\rangle$ [blue dash–dotted line in (d)].
(f) Maximum cross power extracted from each trace in (d) as a function of $\theta_{r}$ [red dash–dotted line in (d)].
In all panels, the experimental data (blue dots) are compared with theoretical fits (solid red lines).}
\end{figure*}

In this study, we utilize the heterodyne detection to process the
time-domain signals, similar to the one used in Ref.~\cite{Bozyigit_NatPhys_2011,Hoi_PRL_2012,Lang_NatPhys_2013,Peng_NatCommun_2016,Zhou_PhysRevAppl_2020}. The
different initial states of qubit are prepared by applying excitation
signals of varying intensities. The qubit states are obtained by measuring
the quadrature amplitudes of spontaneous emission, which are amplified
through a phase insensitive amplifier chain mentioned above. We use
two channels, $a$ and $b$, and add the TWPA to mitigate the influence
of uncorrelated noise in each channel~\cite{Peng_NatCommun_2016,Zhou_PhysRevAppl_2020}.
The expectation value $\langle a(t)\rangle$
is determined through ensemble averaging over $10^{7}$ trials to
obtain satisfactory signal-to-noise ratio.

The excitation is implemented using a Gaussian pulse of the form $A\,\exp[-t^{2}/(2\sigma^{2})]$, with a standard deviation of $\sigma = 4\,\mathrm{ns}$ and a tunable amplitude $A$. This pulse is applied through the qubit’s XY control line to prepare the desired initial states.
Figs.~\ref{fig:Rabi osillation}(a) and \ref{fig:Rabi osillation}(d) show the time evolution of the real part of the emission quadrature amplitude, $\langle S_{a}(t)\rangle \propto \langle a(t)\rangle$, and the cross-power, $\langle S_{b}^{*}(t) S_{a}(t)\rangle \propto \langle a^{\dagger}(t)a(t)\rangle$, respectively. The measurements are performed with the qubit initialized in the state $\cos(\theta_{r}/2)\vert0\rangle+\sin(\theta_{r}/2)\vert1\rangle$, where $\theta_r$ denotes the Rabi angle in the Rabi oscillation sequence. Each trace is obtained from $10^{7}$ ensemble averages~\cite{daSilva_PhysRevA_2010,Bozyigit_NatPhys_2011,Hoi_PRL_2012,Lang_NatPhys_2013,Peng_NatCommun_2016,Zhou_PhysRevAppl_2020}.
We find the quadrature amplitude $\langle a(t)\rangle$ is excellent
agreement with $\sin(\theta_{r})/2$, while power $\langle a^{\dagger}(t)a(t)\rangle$
shows $\sin^{2}(\theta_{r}/2)$. When Rabi angle $\theta_{r}=\pi$
pulse level, the qubit population almost stay on the state $\vert1\rangle$,
corresponding to the single-photon emission case. Consequently, the
cross power $\langle a^{\dagger}(t)a(t)\rangle$ is maximal, while
the quadrature amplitude $\langle a(t)\rangle\approx 0$ because the
phase of the states is uncertain. It should be noted that we adjust
the global phase and set the imaginary part of $\langle a(t)\rangle$
to tend toward zero ~\cite{Bozyigit_NatPhys_2011}. 
The qubit frequency
is not chosen at the sweat point, as it is limited by the output frequency of the AWG and the gain bandwidth of the TWPA. The state preparation fidelity is approximately
0.91, based on calibrating the gain coefficient and comparing the
measured single-photon energy , which is very close to the theoretical
value approximately 0.92~\cite{Zhou_PhysRevAppl_2020,Peng_NatCommun_2016,Lu_NpjQuantumInf_2021}.
Considering the eigen efficiency $\eta$, the total efficiency for
generating a single photon is estimated to be approximately 0.66.
The total efficiency will be improved through sample iterations and
optimization electromagnetic environment.

\begin{figure*}
\includegraphics[width=0.7\textwidth]{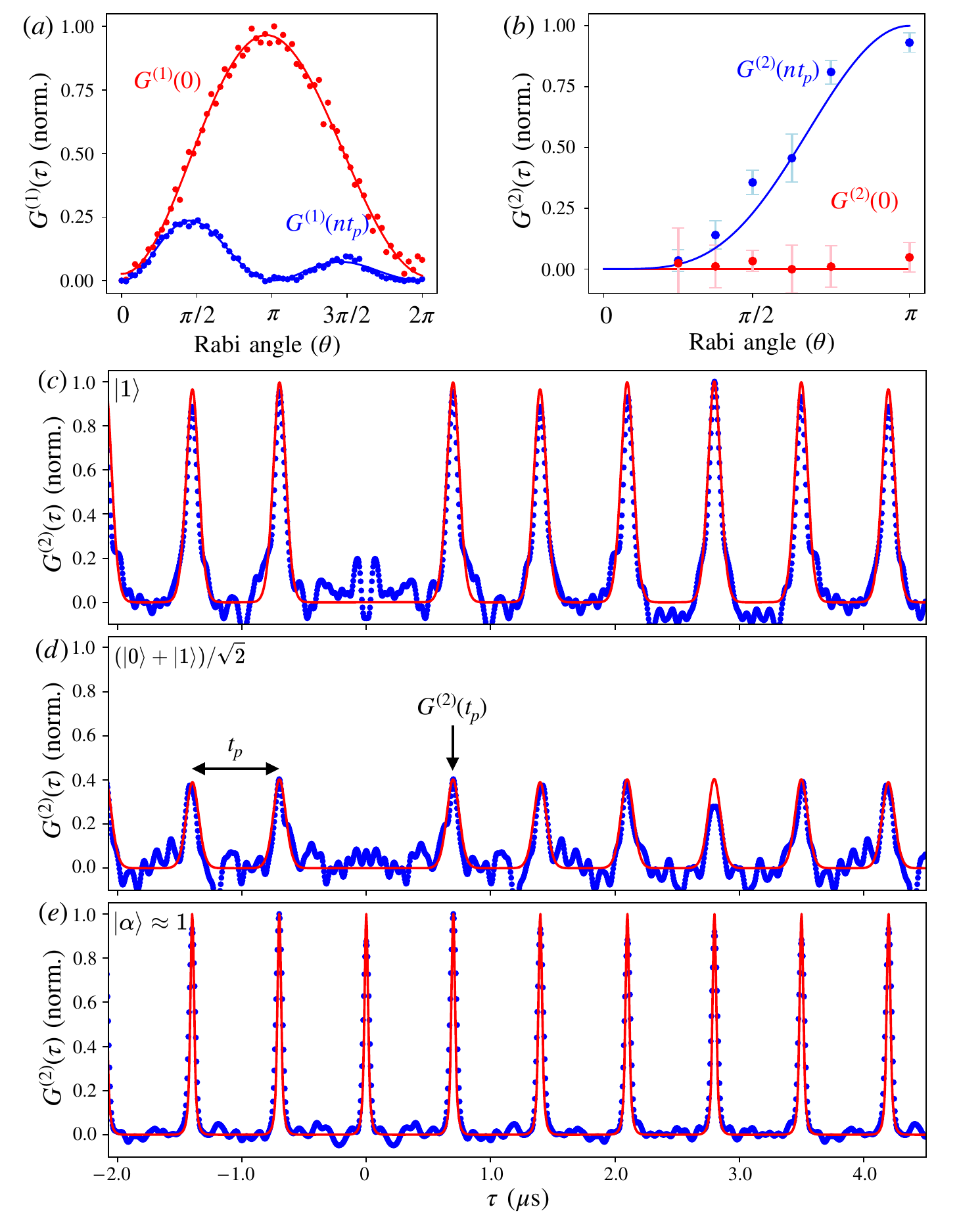}

\caption{\label{fig:corrlation function}Correlation function measurements.
(a) Time dependence of the first-order correlation function $G^{(1)}(\tau)$ evaluated at the central peak ($\tau = 0$) and the side peaks ($\tau = n t_{p}$) as a function of the qubit preparation Rabi angle $\theta_{r}$.
(b) Time dependence of the second-order correlation function $G^{(2)}(\tau)$ at $\tau = 0$ and $\tau = n t_{p}$ versus $\theta_{r}$. Blue and red error bars indicate the standard deviations of the mean values of $G^{(2)}(0)$ and $G^{(2)}(n t_{p})$, respectively.
(c) Measured $G^{(2)}(\tau)$ for the single-photon Fock state $\vert 1\rangle$.
(d) Measured $G^{(2)}(\tau)$ for the superposition state $(\vert 0\rangle + \vert 1\rangle)/\sqrt{2}$.
(e) Measured $G^{(2)}(\tau)$ for a coherent state with amplitude $\vert \alpha \vert \approx 1$.
In all panels, the experimental data (blue dots) are compared with theoretical fits (solid red lines).}
\end{figure*}

To confirm that the emission is indeed of single photons, we utilize
the correlation function of the emitted single photon state using the Hanbury
Brown and Twiss (HBT) setup~\cite{HanburyBrown_nature_1956}. 
We set a control period of 1.6 $\mu s$ for each qubit operation, during which the qubit is excited and then returns to its ground state before being excited again. The effective signal duration is 1.4 $\mu s$, and the correlation measurements reconstruct a train of 8 single-photon pulses with a 700 ns separation between adjacent pulses.
The transmon qubit decays to its ground state before being excited again.
The emitted photons are split into two channels by the hybrid coupler
and subsequently amplified by  the linear amplifier chain. Then, the two-channel signals are down-converted to intermediate frequency (IF) signals at 50 MHz and digitized by ADCs. These signals are then processed by the CPU and GPU to calculate the time-domain correlations between the two quadrature amplitudes,  $S_{a}(t)$  and $S_{b}(t)$, which contain both qubit and noise information. The same method is applied to calculate the first-order cross-correlation function~\cite{daSilva_PhysRevA_2010,Bozyigit_NatPhys_2011,Lang_NatPhys_2013,Peng_NatCommun_2016,Zhou_PhysRevAppl_2020}
\begin{equation}
\Gamma^{(1)}(\tau)=\int\!\mathrm{d}t\langle S_{a}^{*}(t)S_{b}(t+\tau)\rangle,\label{eq:1o_cor_func}
\end{equation}
and the second-order auto-correlation of the cross power
\begin{equation}
\Gamma^{(2)}(\tau)=\int\!\mathrm{d}t\langle S_{a}^{*}(t)S_{a}^{*}(t+\tau)S_{b}(t+\tau)S_{b}(t)\rangle.\label{eq:2_o_cor_func}
\end{equation}
Here, we consider the noise trace follows with each signal trace
, and we calculate and average the traces using the same method to obtain the correlation
function, thereby reducing the noise background influence. Then, the correlation
function of the emitted photons can be expressed as
\begin{align}
G^{(1)}(\tau) & \propto\Gamma^{(1)}(\tau)-\Gamma_{bg}^{(1)}(\tau),\\
G^{(2)}(\tau) & \propto\Gamma^{(2)}(\tau)-\Gamma_{bg}^{(2)}(\tau).
\end{align}
Fig. \ref{fig:corrlation function}(a) shows the dependence of the center
peak $G^{(1)}(0)$ and the side peak $G^{(1)}(nt_{p})$ on the different
Rabi angle $\theta_{r}$, averaged over $10^{7}$ times of 8-photon
sequences with different state preparations. For the single photon
state $\vert1\rangle$ ($\theta_{r}=\pi$), the correlation function
$G^{(1)}(0)$ reaches its maximum and $G^{(1)}(nt_{p})$ vanishes.
The center peak $G^{(1)}(0)\propto\langle a^{\dagger}a\rangle\propto\sin^{2}(\theta_{r}/2)$
measures the average number of emitted photons, while the side peak
$G^{(1)}(nt_{p})\propto\langle a^{\dagger}\rangle\langle a\rangle\propto\sin^{2}(\theta_{r})/4$
indicates the photons generated in different pulses are uncorrelated.
The damping of the oscillations primarily due to the decoherence.

By using the first-order correlation measurements in Fig.~5(a) as a reference for the signal strength, we can more accurately calibrate the $\pi$-pulse amplitude, thereby preparing a nearly pure single-photon signal. At the same time, the second-order correlation function of this state exhibits clear antibunching characteristics. In Fig.~5(c), we present the second-order correlation measurement results of the emitted single photons from the source~\cite{daSilva_PhysRevA_2010,Bozyigit_NatPhys_2011,Hoi_PRL_2012,Lang_NatPhys_2013,Peng_NatCommun_2016,Zhou_PhysRevAppl_2020}. The antibunching effect at zero time delay, $\tau = 0$, confirms the microwave single-photon of the emitted radiation. Since second-order correlation measurements involve higher-order statistical moments, they require a large number of repetitions to obtain reliable statistics. In this experiment, each measurement point in the second-order correlation function was acquired by averaging over nearly $10^{9}$ trials, with data processing carried out on GPU, as mentioned previously. With the present acquisition and processing pipeline, a full second-order correlation measurement can be completed in about $42$ minutes, and further optimization of the single-shot trigger period is expected to reduce this measurement time even further.

To compare the second-order correlations of different quantum states, we measure the correlation functions for several representative input states. In Fig.~5(d), we prepare an equal-weight superposition state $(\vert0\rangle+\vert1\rangle)/\sqrt{2}$ and observe antibunching in a similar manner. We also show the dependence of the second-order correlation functions $G^{2}(0)$ and $G^{2}(\tau)$ on the prepared qubit state, finding excellent agreement with the theoretical predictions. For the second-order correlation of the resulting coherent state $\vert\alpha\approx1\rangle$, as shown in Fig.~5(e), no antibunching is observed, consistent with the expected behavior of a coherent state.

\section{SUMMARY}

In this work, we have fabricated a microwave single-photon source based on tantalum thin films and demonstrated its single-photon character through frequency-domain, time-domain, and correlation measurements. The performance of this tantalum-based source is found to be comparable to that of previously realized aluminum-based devices. In addition, tantalum offers a higher critical temperature and more robust thermal cycling behavior, suggesting that tantalum-based microwave single-photon sources can alleviate some of the limitations associated with earlier material platforms, even though the overall efficiency of the present device still leaves room for improvement. Further optimization of the fabrication process and device design is expected to enhance the emission efficiency and coherence properties.

We have also integrated a TWPA as the first-stage amplifier in the detection chain, which significantly enhances the signal-to-noise ratio and the measurement efficiency. This improvement allows us to complete a full measurement of the second-order cross-correlation function within $40$ minutes, representing more than 50 times acceleration in overall measurement speed compared with previous single-photon characterization experiments performed without a TWPA~\cite{Peng_NatCommun_2016,Zhou_PhysRevAppl_2020}. As a result, the single-photon statistics of the source can now be verified much faster and more reliably. Based on these results, tantalum-based microwave single-photon sources, combined with tunable couplers for active control of the qubit emission rate and photon-waveform shaping, provide a promising platform for future Hong–Ou–Mandel interferometer~\cite{Hong_PRL_1987,Maunz_NP_2007,Woolley_NJP_2013} and Mach–Zehnder interferometer~\cite{Santori_nature_2002}, as well as for long-distance quantum communication with flying microwave photons~\cite{Zhong_NatPhys_2019,Zhong_Nature_2021,Qiu_SciBull_2025,gong_arxiv_2024}.
\begin{acknowledgments}
This work was supported by the National Natural Science Foundation of China (NSFC) with Grants No.~92576206 and 92365209.
\end{acknowledgments}


\bibliography{apssamp}

@article{Koch_PRA_2007,
  author    = {J. Koch and T. M. Yu and J. Gambetta and A. A. Houck and D. I. Schuster and J. Majer and A. Blais and M. H. Devoret and S. M. Girvin and R. J. Schoelkopf},
  title     = {{Charge-Insensitive Qubit Design Derived from the Cooper Pair Box}},
  journal   = {Phys. Rev. A},
  year      = {2007},
  volume    = {76},
  number    = {4},
  pages     = {042319},
  doi       = {10.1103/PhysRevA.76.042319}
}

@article{You_PRB_2007,
  title = {Low-decoherence flux qubit},
  author = {You, J. Q. and Hu, Xuedong and Ashhab, S. and Nori, Franco},
  journal = {Phys. Rev. B},
  volume = {75},
  issue = {14},
  pages = {140515},
  numpages = {4},
  year = {2007},
  month = {Apr},
  publisher = {American Physical Society},
  doi = {10.1103/PhysRevB.75.140515},
  url = {https://link.aps.org/doi/10.1103/PhysRevB.75.140515}
}

@article{Gabelli_PRL_2004,
  author    = {J. Gabelli and L.-H. Reydellet and G. F\`eve and J.-M. Berroir and B. Pla\c{c}ais and P. Roche and D. C. Glattli},
  title     = {{Hanbury Brown--Twiss Correlations to Probe the Population Statistics of GHz Photons Emitted by Conductors}},
  journal   = {Phys. Rev. Lett.},
  year      = {2004},
  volume    = {93},
  number    = {5},
  pages     = {056801},
  doi       = {10.1103/PhysRevLett.93.056801}
}

@article{Astafiev_Science_2010,
  author    = {O. Astafiev and A. M. Zagoskin and A. A. Abdumalikov and Y. A. Pashkin and T. Yamamoto and K. Inomata and Y. Nakamura and J. S. Tsai},
  title     = {{Resonance Fluorescence of a Single Artificial Atom}},
  journal   = {Science},
  year      = {2010},
  volume    = {327},
  number    = {5967},
  pages     = {840--843},
  doi       = {10.1126/science.1181918}
}

@article{Besse_PRX_2018,
  author    = {J.-C. Besse and S. Gasparinetti and M. C. Collodo and T. Walter and P. Kurpiers and M. Pechal and C. Eichler and A. Wallraff},
  title     = {{Single-Shot Quantum Nondemolition Detection of Individual Itinerant Microwave Photons}},
  journal   = {Phys. Rev. X},
  year      = {2018},
  volume    = {8},
  number    = {2},
  pages     = {021003},
  doi       = {10.1103/PhysRevX.8.021003}
}

@article{Kono_NatPhys_2018,
  author    = {S. Kono and K. Koshino and Y. Tabuchi and A. Noguchi and Y. Nakamura},
  title     = {{Quantum Non-Demolition Detection of an Itinerant Microwave Photon}},
  journal   = {Nat. Phys.},
  year      = {2018},
  volume    = {14},
  number    = {6},
  pages     = {546--549},
  doi       = {10.1038/s41567-018-0066-3}
}

@article{Inomata_NatCommun_2016,
  author = {K. Inomata and Z. Lin and K. Koshino and W. D. Oliver and J. S. Tsai and T. Yamamoto and Y. Nakamura},
  title = {{Single Microwave-Photon Detector Using an Artificial $\Lambda$-Type Three-Level System}},
  journal = {Nat. Commun.},
  volume = {7},
  pages = {12303},
  year = {2016},
  doi = {10.1038/ncomms12303}
}

@article{daSilva_PhysRevA_2010,
  author = {M. P. da Silva and D. Bozyigit and A. Wallraff and A. Blais},
  title = {{Schemes for the Observation of Photon Correlation Functions in Circuit QED with Linear Detectors}},
  journal = {Phys. Rev. A},
  volume = {82},
  number = {4},
  pages = {043804},
  year = {2010},
  doi = {10.1103/PhysRevA.82.043804}
}

@article{Lang_NatPhys_2013,
  author = {C. Lang and C. Eichler and L. Steffen and J. M. Fink and M. J. Woolley and A. Blais and A. Wallraff},
  title = {{Correlations, Indistinguishability and Entanglement in Hong--Ou--Mandel Experiments at Microwave Frequencies}},
  journal = {Nat. Phys.},
  volume = {9},
  number = {6},
  pages = {345--348},
  year = {2013},
  doi = {10.1038/nphys2612}
}

@article{CastellanosBeltran_APL_2007,
  author  = {Castellanos-Beltran, M. A. and Lehnert, K. W.},
  title   = {Widely tunable parametric amplifier based on a superconducting quantum interference device array resonator},
  journal = {Appl. Phys. Lett.},
  volume  = {91},
  number  = {8},
  pages   = {083509},
  year    = {2007},
  doi     = {10.1063/1.2773988}
}

@article{Yamamoto_APL_2008,
  author  = {Yamamoto, T. and Inomata, K. and Watanabe, M. and Matsuba, K. and Miyazaki, T. and Oliver, W. D. and Nakamura, Y. and Tsai, J. S.},
  title   = {{Flux-driven Josephson parametric amplifier}},
  journal = {Appl. Phys. Lett.},
  volume  = {93},
  number  = {4},
  pages   = {042510},
  year    = {2008},
  doi     = {10.1063/1.2964182}
}

@article{Aumentado_IEEE_2020,
  author  = {Aumentado, Jose},
  title   = {{Superconducting Parametric Amplifiers: The State of the Art in Josephson Parametric Amplifiers}},
  journal = {IEEE Microw. Mag.},
  volume  = {21},
  number  = {8},
  pages   = {45--59},
  year    = {2020},
  doi     = {10.1109/MMM.2020.2993476}
}

@article{HoEom_NP_2012,
  author  = {Ho Eom, Byeong and Day, Peter K. and LeDuc, Henry G. and Zmuidzinas, Jonas},
  title   = {{A wideband, low-noise superconducting amplifier with high dynamic range}},
  journal = {Nat. Phys.},
  volume  = {8},
  number  = {8},
  pages   = {623--627},
  year    = {2012},
  doi     = {10.1038/nphys2356}
}

@article{Macklin_Science_2015,
  author = {C. Macklin and K. O'Brien and D. Hover and M. E. Schwartz and V. Bolkhovsky and X. Zhang and W. D. Oliver and I. Siddiqi},
  title = {{A near--quantum-limited Josephson traveling-wave parametric amplifier}},
  journal = {Science},
  volume = {350},
  number = {6258},
  pages = {307--310},
  year = {2015},
  doi = {10.1126/science.aaa8525}
}

@article{Brien_PhysRevLett_2014,
  title = {{Resonant Phase Matching of Josephson Junction Traveling Wave Parametric Amplifiers}},
  author = {O'Brien, Kevin and Macklin, Chris and Siddiqi, Irfan and Zhang, Xiang},
  journal = {Phys. Rev. Lett.},
  volume = {113},
  issue = {15},
  pages = {157001},
  numpages = {5},
  year = {2014},
  month = {Oct},
  publisher = {American Physical Society},
  doi = {10.1103/PhysRevLett.113.157001},
  url = {https://link.aps.org/doi/10.1103/PhysRevLett.113.157001}
}

@article{Bockstiegel_JLTP_2014,
	author = {Bockstiegel, C. and Gao, J. and Vissers, M. R. and Sandberg, M. and Chaudhuri, S. and Sanders, A. and Vale, L. R. and Irwin, K. D. and Pappas, D. P.},
	journal = {J. Low Temp. Phys.},
	number = {3},
	pages = {476--482},
	title = {{Development of a Broadband NbTiN Traveling Wave Parametric Amplifier for MKID Readout}},
	volume = {176},
	year = {2014}}

@article{Esposito_APL_2021,
  author = {Esposito, Martina and Ranadive, Arpit and Planat, Luca and Roch, Nicolas},
  title = {{Perspective on traveling wave microwave parametric amplifiers}},
  journal = {Appl. Phys. Lett.},
  volume = {119},
  number = {12},
  pages = {120501},
  year = {2021},
  doi = {10.1063/5.0064892}
}

@article{Brown_Nature_1956,
  author = {R. H. Brown and R. Q. Twiss},
  title = {{Correlation between Photons in two Coherent Beams of Light}},
  journal = {Nature},
  volume = {177},
  number = {4497},
  pages = {27--29},
  year = {1956},
  doi = {10.1038/177027a0}
}

@article{Lu_NpjQuantumInf_2021,
  author = {Y. Lu and A. Bengtsson and J. J. Burnett and B. Suri and S. R. Sathyamoorthy and H. R. Nilsson and M. Scigliuzzo and J. Bylander and G. Johansson and P. Delsing},
  title = {{Quantum Efficiency, Purity and Stability of a Tunable, Narrowband Microwave Single-Photon Source}},
  journal = {npj Quantum Inf.},
  volume = {7},
  number = {1},
  pages = {140},
  year = {2021},
  doi = {10.1038/s41534-021-00480-5}
}

@article{Bozyigit_NatPhys_2011,
  author = {D. Bozyigit and C. Lang and L. Steffen and J. M. Fink and C. Eichler and M. Baur and R. Bianchetti and P. J. Leek and S. Filipp and M. P. da Silva and A. Blais and A. Wallraff},
  title = {{Antibunching of Microwave-Frequency Photons Observed in Correlation Measurements Using Linear Detectors}},
  journal = {Nat. Phys.},
  volume = {7},
  number = {2},
  pages = {154--158},
  year = {2011},
  doi = {10.1038/nphys1845}
}

@article{Hoi_PRL_2012,
  title = {{Generation of Nonclassical Microwave States Using an Artificial Atom in 1D Open Space}},
  author = {Hoi, Io-Chun and Palomaki, Tauno and Lindkvist, Joel and Johansson, G\"oran and Delsing, Per and Wilson, C. M.},
  journal = {Phys. Rev. Lett.},
  volume = {108},
  issue = {26},
  pages = {263601},
  numpages = {5},
  year = {2012},
  month = {Jun},
  publisher = {American Physical Society},
  doi = {10.1103/PhysRevLett.108.263601},
  url = {https://link.aps.org/doi/10.1103/PhysRevLett.108.263601}
}

@article{Wang_npjQI_2022,
  author = {C. Wang and X. Li and H. Xu and Z. Li and J. Wang and Z. Yang and Z. Mi and X. Liang and T. Su and C. Yang and G. Wang and W. Wang and Y. Li and M. Chen and C. Li and K. Linghu and J. Han and Y. Zhang and Y. Feng and Y. Song and T. Ma and J. Zhang and R. Wang and P. Zhao and W. Liu and G. Xue and Y. Jin and H. Yu},
  title = {{Towards Practical Quantum Computers: Transmon Qubit with a Lifetime Approaching 0.5 Milliseconds}},
  journal = {npj Quantum Inf.},
  volume = {8},
  number = {1},
  pages = {3},
  year = {2022},
  doi = {10.1038/s41534-021-00510-2}
}

@article{Russo_SST_2005,
  author = {R. Russo and L. Catani and A. Cianchi and S. Tazzari and J. Langner},
  title = {{High Quality Superconducting Niobium Films Produced by an Ultra-High Vacuum Cathodic Arc}},
  journal = {Supercond. Sci. Technol.},
  volume = {18},
  number = {7},
  pages = {L41--L44},
  year = {2005},
  doi = {10.1088/0953-2048/18/7/L01}
}

@article{Valderrama_AIPCP_2012,
  author = {E. F. Valderrama and C. James and M. Krishnan and X. Zhao and L. Phillips and C. Reece and K. Seo},
  title = {{High-{RRR} Thin-Films of {Nb} Produced Using Energetic Condensation from a Coaxial, Rotating Vacuum {ARC} Plasma ({CEDTM})}},
  journal = {AIP Conf. Proc.},
  volume = {1434},
  number = {1},
  pages = {953--960},
  year = {2012},
  doi = {10.1063/1.4707012}
}

@article{Premkumar_CM_2021,
  author = {A. Premkumar and C. Weiland and S. Hwang and B. J{\"a}ck and A. P. M. Place and I. Waluyo and A. Hunt and V. Bisogni and J. Pelliciari and A. Barbour and M. S. Miller and P. Russo and F. Camino and K. Kisslinger and X. Tong and M. S. Hybertsen and A. A. Houck and I. Jarrige},
  title = {{Microscopic Relaxation Channels in Materials for Superconducting Qubits}},
  journal = {Commun. Mater.},
  volume = {2},
  number = {1},
  pages = {72},
  year = {2021},
  doi = {10.1038/s43246-021-00168-8}
}

@article{Ding_JVSTB_2024,
  author = {Z. Ding and B. Zhou and T. Wang and L. Yang and Y. Wu and X. Cai and K. Xiong and J. Feng},
  title = {{Stable and Low Loss Oxide Layer on $\alpha$-Ta (110) Film for Superconducting Qubits}},
  journal = {J. Vac. Sci. Technol. B},
  volume = {42},
  number = {2},
  pages = {022209},
  year = {2024},
  doi = {10.1116/6.0003368}
}

@misc{gong_arxiv_2024,
      title={Tunable quantum router with giant atoms, implementing quantum gates, teleportation, non-reciprocity, and circulators}, 
      author={Rui-Yang Gong and Zi-Yu He and Cheng-He Yu and Ge-Fei Zhang and Franco Nori and Ze-Liang Xiang},
      year={2024},
      eprint={2411.19307},
      archivePrefix={arXiv},
      primaryClass={quant-ph},
      url={https://arxiv.org/abs/2411.19307}, 
}

@article{Zhou_JJAP_2023,
doi = {10.35848/1347-4065/acfde6},
url = {https://doi.org/10.35848/1347-4065/acfde6},
year = {2023},
month = {oct},
publisher = {IOP Publishing},
volume = {62},
number = {10},
pages = {100901},
author = {Zhou, Boyi and Yang, Lina and Wang, Tao and Wang, Yu and Ding, Zengqian and Wu, Yanfu and Xiong, Kanglin and Feng, Jiagui},
title = {{E}pitaxial $\alpha$-{T}a (110) film on a-plane sapphire substrate for superconducting qubits on wafer scale},
journal = {Jpn. J. Appl. Phys.},

}

@article{Zheng_SR_2023,
  author = {Y. Zheng and S. Li and Z. Ding and K. Xiong and J. Feng and H. Yang},
  title = {{Fabrication of {Al}/{AlOx}/{Al} Junctions with High Uniformity and Stability on Sapphire Substrates}},
  journal = {Sci. Rep.},
  volume = {13},
  number = {1},
  pages = {11874},
  year = {2023},
  doi = {10.1038/s41598-023-39052-2}
}

@article{Zhou_PhysRevAppl_2020,
  author = {Y. Zhou and Z. Peng and Y. Horiuchi and O. V. Astafiev and J. S. Tsai},
  title = {{Tunable Microwave Single-Photon Source Based on Transmon Qubit with High Efficiency}},
  journal = {Phys. Rev. Appl.},
  year = {2020},
  volume = {13},
  number = {3},
  pages = {034007},
  doi = {10.1103/PhysRevApplied.13.034007}
}

@article{Peng_NatCommun_2016,
  author    = {Z. H. Peng and S. E. de Graaf and J. S. Tsai and O. V. Astafiev},
  title     = {{Tuneable On-Demand Single-Photon Source in the Microwave Range}},
  journal   = {Nat. Commun.},
  year      = {2016},
  volume    = {7},
  pages     = {12588},
  doi       = {10.1038/ncomms12588}
}

@article{Axline_NatPhy_2018,
  author = {Axline, Christopher J. and Burkhart, Luke D. and Pfaff, Wolfgang and Zhang, Mengzhen and Chou, Kevin and Campagne-Ibarcq, Philippe and Reinhold, Philip and Frunzio, Luigi and Girvin, S. M. and Jiang, Liang and Devoret, M. H. and Schoelkopf, R. J.},
  title = {{On-demand quantum state transfer and entanglement between remote microwave cavity memories}},
  journal = {Nat. Phys.},
  volume = {14},
  number = {7},
  pages = {705--710},
  year = {2018}
}

@article{Kurpiers_Nature_2018,
  author = {P. Kurpiers and P. Magnard and T. Walter and B. Royer and M. Pechal and J. Heinsoo and Y. Salathé and A. Akin and S. Storz and J.-C. Besse and S. Gasparinetti and A. Blais and A. Wallraff},
  title = {{Deterministic Quantum State Transfer and Remote Entanglement Using Microwave Photons}},
  journal = {Nature},
  year = {2018},
  volume = {558},
  pages = {264--267},
  doi = {10.1038/s41586-018-0195-y}
}

@article{Zhong_NatPhys_2019,
	author = {Zhong, Y. P. and Chang, H. -S. and Satzinger, K. J. and Chou, M. -H. and Bienfait, A. and Conner, C. R. and Dumur, {\'E}. and Grebel, J. and Peairs, G. A. and Povey, R. G. and Schuster, D. I. and Cleland, A. N.},
	journal = {Nat. Phys.},
	number = {8},
	pages = {741--744},
	title = {{Violating Bell's inequality with remotely connected superconducting qubits}},
	volume = {15},
	year = {2019}}

@article{Zhong_Nature_2021,
	author = {Zhong, Youpeng and Chang, Hung-Shen and Bienfait, Audrey and Dumur, {\'E}tienne and Chou, Ming-Han and Conner, Christopher R. and Grebel, Joel and Povey, Rhys G. and Yan, Haoxiong and Schuster, David I. and Cleland, Andrew N.},
	journal = {Nature},
	number = {7847},
	pages = {571--575},
	title = {{Deterministic multi-qubit entanglement in a quantum network}},
	volume = {590},
	year = {2021}}

@article{Qiu_SciBull_2025,
	author = {Qiu, Jiawei and Liu, Yang and Hu, Ling and Wu, Yukai and Niu, Jingjing and Zhang, Libo and Huang, Wenhui and Chen, Yuanzhen and Li, Jian and Liu, Song and Zhong, Youpeng and Duan, Luming and Yu, Dapeng},
	journal = {Sci. Bull.},
	number = {3},
	pages = {351--358},
	title = {{Deterministic quantum state and gate teleportation between distant superconducting chips}},
	volume = {70},
	year = {2025}}

@article{Kjaergaard_ARCMP_2020,
  author  = {Kjaergaard, Morten and Schwartz, Mollie E. and Braumüller, Jochen and Krantz, Philip and Wang, Joel I.-J. and Gustavsson, Simon and Oliver, William D.},
  title   = {{Superconducting Qubits: Current State of Play}},
  journal = {Annu. Rev. Condens. Matter Phys.},
  volume  = {11},
  pages   = {369--395},
  year    = {2020},
  doi     = {10.1146/annurev-conmatphys-031119-050605}
}

@article{Azuma_RMP_2023,
  title = {{Quantum repeaters: From quantum networks to the quantum internet}},
  author = {Azuma, Koji and Economou, Sophia E. and Elkouss, David and Hilaire, Paul and Jiang, Liang and Lo, Hoi-Kwong and Tzitrin, Ilan},
  journal = {Rev. Mod. Phys.},
  volume = {95},
  issue = {4},
  pages = {045006},
  numpages = {66},
  year = {2023},
  month = {Dec},
  publisher = {American Physical Society},
  doi = {10.1103/RevModPhys.95.045006},
  url = {https://link.aps.org/doi/10.1103/RevModPhys.95.045006}
}

@article{Jiang_nphQI_2021,
  author  = {Rozp{\k{e}}dek, Filip and Noh, Kyungjoo and Xu, Qian and Guha, Saikat and Jiang, Liang},
  title   = {{Quantum repeaters based on concatenated bosonic and discrete-variable quantum codes}},
  journal = {npj Quantum Inf.},
  volume  = {7},
  pages   = {102},
  year    = {2021}
}

@article{Kimble_nature_2008,
	author = {Kimble, H.  J. },
	journal = {Nature},
	number = {7198},
	pages = {1023--1030},
	title = {The quantum internet},
	volume = {453},
	year = {2008}}

@article{Couteau_NRP_2023,
  author = {Couteau, Christophe and Barz, Stefanie and Durt, Thomas and Gerrits, Thomas and Huwer, Jan and Prevedel, Robert and Rarity, John and Shields, Andrew and Weihs, Gregor},
  title = {{Applications of single photons to quantum communication and computing}},
  journal = {Nat. Rev. Phys.},
  volume = {5},
  number = {6},
  pages = {326--338},
  year = {2023}
}

@article{Blais_RevModPhys_2021,
  title = {{Circuit quantum electrodynamics}},
  author = {Blais, Alexandre and Grimsmo, Arne L. and Girvin, S. M. and Wallraff, Andreas},
  journal = {Rev. Mod. Phys.},
  volume = {93},
  issue = {2},
  pages = {025005},
  numpages = {72},
  year = {2021},
  month = {May},
  publisher = {American Physical Society},
  doi = {10.1103/RevModPhys.93.025005},
  url = {https://link.aps.org/doi/10.1103/RevModPhys.93.025005}
}

@article{Cirac_PRL_1997,
  title = {{Quantum State Transfer and Entanglement Distribution among Distant Nodes in a Quantum Network}},
  author = {Cirac, J. I. and Zoller, P. and Kimble, H. J. and Mabuchi, H.},
  journal = {Phys. Rev. Lett.},
  volume = {78},
  issue = {16},
  pages = {3221--3224},
  numpages = {0},
  year = {1997},
  month = {Apr},
  publisher = {American Physical Society},
  doi = {10.1103/PhysRevLett.78.3221},
  url = {https://link.aps.org/doi/10.1103/PhysRevLett.78.3221}
}

@article{Zoller_EPJD_2005,
  author = {Zoller, P. and Beth, Th. and Binosi, D. and Blatt, R. and Briegel, H. and Bruss, D. and Calarco, T. and Cirac, J. I. and Deutsch, D. and Eisert, J. and Ekert, A. and Fabre, C. and Gisin, N. and Grangiere, P. and Grassl, M. and Haroche, S. and Imamoglu, A. and Karlson, A. and Kempe, J. and Kouwenhoven, L. and Kroell, S. and Leuchs, G. and Lewenstein, M. and Loss, D. and Lutkenhaus, N. and Massar, S. and Mooij, J. E. and Plenio, M. B. and Polzik, E. and Popescu, S. and Rempe, G. and Sergienko, A. and Suter, D. and Twamley, J. and Wendin, G. and Werner, R. and Winter, A. and Wrachtrup, J. and Zeilinger, A.},
  title = {{Quantum information processing and communication}},
  journal = {Eur. Phys. J. D},
  volume = {36},
  number = {2},
  pages = {203--228},
  year = {2005}
}

@article{Place_NC_2021,
  author  = {Place, Alexander P. M. and Rodgers, Lila V. H. and Mundada, Pranav and Smitham, Basil M. and Fitzpatrick, Mattias and Leng, Zhaoqi and Premkumar, Anjali and Bryon, Jacob and Vrajitoarea, Andrei and Sussman, Sara and Cheng, Guangming and Madhavan, Trisha and Babla, Harshvardhan K. and Le, Xuan Hoang and Gang, Youqi and J{\"a}ck, Berthold and Gyenis, Andr{\'a}s and Yao, Nan and Cava, Robert J. and de Leon, Nathalie P. and Houck, Andrew A.},
  title   = {New material platform for superconducting transmon qubits with coherence times exceeding 0.3 milliseconds},
  journal = {Nat. Commun.},
  volume  = {12},
  pages   = {1779},
  year    = {2021}
}

@article{Lubsanov_SST_2022,
doi = {10.1088/1361-6668/ac8a24},
url = {https://doi.org/10.1088/1361-6668/ac8a24},
year = {2022},
month = {sep},
publisher = {IOP Publishing},
volume = {35},
number = {10},
pages = {105013},
author = {Lubsanov, Viktor and Gurtovoi, Vladimir and Semenov, Alexander and Glushkov, Evgenii and Antonov, Vladimir and Astafiev, Oleg},
title = {{Materials for a broadband microwave superconducting single photon detector}},
journal = {Supercond. Sci. Technol.},

}

@article{HanburyBrown_nature_1956,
  author  = {R. Hanbury Brown and R. Q. Twiss},
  title   = {{Correlation between Photons in Two Coherent Beams of Light}},
  journal = {Nature},
  volume  = {177},
  number  = {4497},
  pages   = {27--29},
  year    = {1956},
  doi     = {10.1038/177027a0}
}

@article{Hong_PRL_1987,
  title = {{Measurement of subpicosecond time intervals between two photons by interference}},
  author = {Hong, C. K. and Ou, Z. Y. and Mandel, L.},
  journal = {Phys. Rev. Lett.},
  volume = {59},
  issue = {18},
  pages = {2044--2046},
  numpages = {0},
  year = {1987},
  month = {Nov},
  publisher = {American Physical Society},
  doi = {10.1103/PhysRevLett.59.2044},
  url = {https://link.aps.org/doi/10.1103/PhysRevLett.59.2044}
}

@article{Maunz_NP_2007,
  author  = {P. Maunz and D. L. Moehring and S. Olmschenk and K. C. Younge and D. N. Matsukevich and C. Monroe},
  title   = {{Quantum interference of photon pairs from two remote trapped atomic ions}},
  journal = {Nat. Phys.},
  volume  = {3},
  number  = {8},
  pages   = {538--541},
  year    = {2007},
  doi     = {10.1038/nphys644}
}

@article{Woolley_NJP_2013,
  author  = {M. J. Woolley and C. Lang and C. Eichler and A. Wallraff and A. Blais},
  title   = {Signatures of Hong--Ou--Mandel interference at microwave frequencies},
  journal = {New J. Phys.},
  volume  = {15},
  number  = {10},
  pages   = {105025},
  year    = {2013},
  doi     = {10.1088/1367-2630/15/10/105025}
}

@article{Santori_nature_2002,
  author  = {C. Santori and D. Fattal and J. Vu{\v c}kovi{\'c} and G. S. Solomon and Y. Yamamoto},
  title   = {{Indistinguishable photons from a single-photon device}},
  journal = {Nature},
  volume  = {419},
  number  = {6907},
  pages   = {594--597},
  year    = {2002},
  doi     = {10.1038/nature01086}
}

\end{document}